\documentclass[aps,prl,twocolumn,superscriptaddress]{revtex4}

\usepackage[usenames,dvipsnames]{color}
\usepackage{amssymb}
\usepackage{graphicx}
\usepackage{cancel}
\usepackage{amsmath}
\usepackage{hyperref}

\newcommand{\figfolder}[1]{}

\newcommand{\jfi}{James Franck Institute and Department of Physics, University of Chicago, Chicago, Illinois 60637}

\begin{document}

\title{High contrast qubit interactions using multimode cavity QED}
\author{David C. McKay}
\email[Email:]{dcmckay@uchicago.edu}
\affiliation{\jfi}
\author{Ravi Naik}
\affiliation{\jfi}
\author{Philip Reinhold}
\altaffiliation[Current address:]{Yale Department of Applied Physics, New Haven, CT 06511}
\affiliation{\jfi}
\author{Lev S. Bishop}
\affiliation{Condensed Matter Theory Center, Department of Physics, University of Maryland, College Park, MD 20742, USA}
\affiliation{IBM T.J. Watson Research Center, Yorktown Heights, New York 10598, USA}
\author{David I. Schuster}
\affiliation{\jfi}
\date{October 2014}

\begin{abstract}
We introduce a new multimode cavity QED architecture for superconducting circuits which can be used to implement photonic memories, more efficient Purcell filters, and quantum simulations of photonic materials.  We show that qubit interactions mediated by multimode cavities can have exponentially improved contrast for two qubit gates without sacrificing gate speed.  Using two-qubits coupled via a three-mode cavity system we spectroscopically observe multimode strong couplings up to 102MHz and demonstrate suppressed interactions off-resonance of 10kHz when the qubits are $\approx$600MHz detuned from the cavity resonance. We study Landau-Zener transitions in our multimode systems and demonstrate quasi-adiabatic loading of single photons into the multimode cavity in 25ns. We introduce an adiabatic gate protocol to realize a controlled-Z gate between the qubits in 95ns and create a Bell state with 94.7\% fidelity. This corresponds to an on/off ratio (gate contrast) of 1000.
\end{abstract}

\maketitle

Circuit cavity quantum electrodynamics (cQED) using superconducting resonators and Josephson junction based qubits have demonstrated the essential building blocks of gate based quantum computing and quantum optics\cite{devoret:2013}.  Typically, cQED devices are engineered so that the qubits primarily couple to a single cavity mode, nonetheless, the true multimode nature of these devices is unavoidable.  For example, a multimode treatment is required to correctly understand the Purcell effect\cite{houck:2008}, and in 3D resonators, where the mode density is higher, it is necessary to take into account the full mode structure using the ``black box quantization'' approach to correctly model the device parameters\cite{nigg:2012}. Although these modes are usually treated as a nuisance, if properly utilized, they are a powerful asset. In this letter, we introduce an explicitly multimode QED architecture as a resource to study multimode quantum optics\cite{egger:2013}, as a many-body bosonic system for quantum simulation\cite{zhou:2008,koch:2013}, as a photonic register for quantum memory, and to tailor coherent qubit-qubit interactions. \\
In the context of quantum computing, tailoring qubit interactions is of paramount importance for improving gate contrast. In the past several years much effort has been spent to improve gate fidelities; single qubit coherence times can approach 100$\mu$s \cite{rigetti:2012}, arbitrary rotations in the Bloch sphere are possible with gate fidelities higher than 99.8\%\cite{gustavsson:2013}, and elementary two-qubit gates have attained gate fidelities up to 99.4\%\cite{barends:2014}. There has been rapid progress towards constructing larger circuits to implement quantum algorithms\cite{dicarlo:2010,fedorov:2012,chow:2013,saira:2013,barends:2014}, photonic memories\cite{mariantoni:2011}, and nascent quantum simulation\cite{underwood:2012}. However, as strongly coupled circuits grow larger, issues inevitably arise due to residual cQED couplings.  \\
Several methods have been developed to reduce unwanted interactions, however, they are not without their limitations. To counteract ``always on'' interactions in NMR quantum computing, decoupling pulse sequences have been developed\cite{criger:2012}. These sequences can be applied to JJ qubits, but become onerous as the system size grows larger. Instead, another approach is to develop tunable interactions for high contrast gates, most commonly by coupling JJ qubits through a resonant interaction. In these experiments, interactions are controlled via the detuning from resonance, imposing a tradeoff between gate contrast and speed. Also, expanding beyond two qubits results in spectral crowding, which limits addressability\cite{schutjens:2013} and introduces spurious avoided crossings. While parametric gates\cite{chow:2011} sidestep some of these problems, in both cases the contrast is only linear in detuning. This limits the achievable off-rate since detuning is bounded. Alternatively, we can dynamically tune the coupling by destructive interference between two charge qubits\cite{srinivasan:2011} or by flux tuning a JJ inductive coupler\cite{bialczak:2011,chen:2014}. However, dynamic coupling requires additional junctions, which introduces complexity and a new path for decoherence. \\ 
In this letter, we introduce a new multimode circuit QED architecture where qubits interact through a network of strongly coupled resonators, analogous to a multimode bandpass filter. The multimode architecture enables the off-resonant interactions to be suppressed exponentially in the number of modes (resonators) without any additional active elements. This multimode architecture could also be utilized to filter the qubit noise environment, i.e. a multimode Purcell filter\cite{reed:2010,jeffrey:2014}. To demonstrate the multimode architecture, we construct a circuit with two transmon-type qubits coupled via a three-mode (three-resonator) filter. We perform spectroscopy on our device and confirm the multimode circuit QED model. From spectroscopy, we observe multimode strong coupling when the qubit and filter are on-resonance and suppressed qubit-qubit interactions off-resonance. Next, we measure strong interaction dynamics by quickly tuning the qubit energy into resonance with the filter. We demonstrate fast loading of single photons into the lowest mode of the filter ($\approx 25$ns) and measure a single photon Stark shift greater than 100MHz. Finally, we utilize the state-dependent Stark shift to realize a controlled-Z gate between the qubits in 95ns and create a Bell state with 94.7\% fidelity.\\
\begin{figure*}[htb]
\includegraphics[width=0.95\textwidth]{\figfolder{1}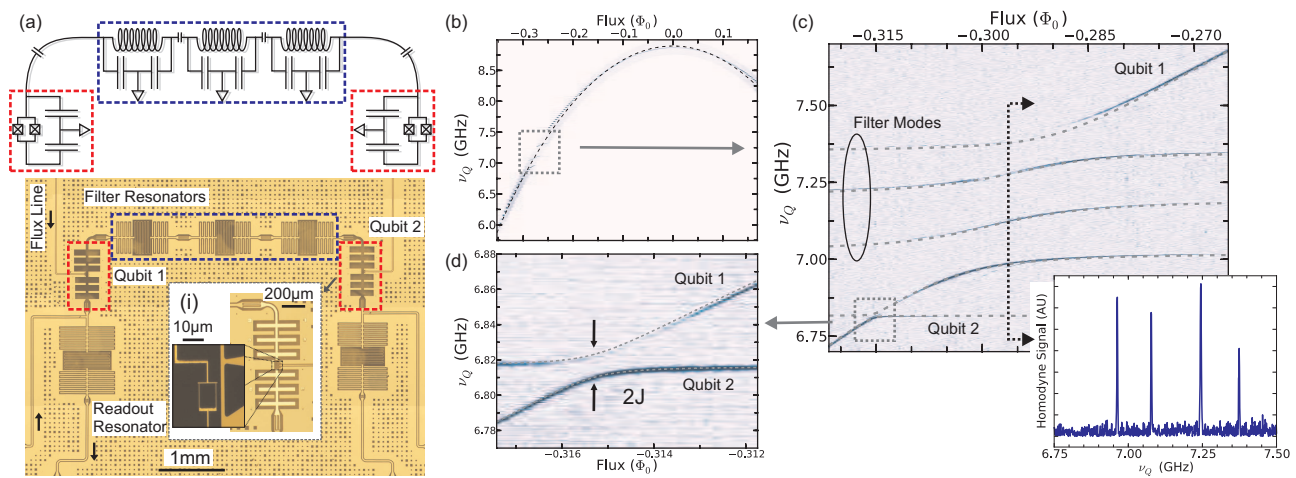}
\caption{(Color Online) {\bf Multimode device schematic and spectroscopy.} (a) Schematic (top) and optical image (bottom) of our 3-resonator cQED device. The schematic shows the three lumped LC resonators of the filter (blue dotted line) that couple two transmon-type qubits (red dotted line). These same features are outlined on the physical circuit. All resonators and large features are Nb (light yellow) etched on sapphire (dark). The qubits have not been deposited in the largest image, but are illustrated after aluminum deposition in the inset (i). Transmission lines from the top-left/top-right allow for fast-flux biasing of the transmon. The qubits are also coupled to readout resonators at $\nu_{1(2)}=4.20 (4.65)$GHz and the qubit state is inferred by measuring the transmission of these resonators. The qubit 1 (qubit 2) lifetime $T_1=$2.36(2.14)$\mu$s and the decay of Ramsey coherence (fit to Gaussian decay $e^{-t^2/2\sigma^2}$) is $\sigma=$312(492)ns. Full fabrication details, qubit properties, instrumentation, and cryogenic setup are given in the supplementary information. (b) Single qubit spectroscopy as the qubit frequency $\nu_Q$ is tuned using the flux line. The dashed line is a fit obtained by diagonalizing the energy levels of the transmon in the charge basis. (c) Spectroscopy of the region where the qubit frequency crosses through the filter modes (dashed box in (b)). The frequency of the other qubit is fixed and below the filter. The dashed lines are the eigenvalues of the Hamiltonian given by Eqn.~\ref{eqn:ham1} using the qubit-filter parameters listed in the main text. The inset is a cross-section of similar spectroscopy data demonstrating multimode strong coupling. (d) Spectroscopy of the qubit-qubit avoided crossing (dashed box in (c)). The minimum frequency separation gives twice the exchange frequency 2J (Eqn.~\ref{eqn:ham2}). In (b),(c) and (d) flux (in $\Phi_0$) is obtained from experimental units as a fit parameter.   \label{fig:1}}
\end{figure*}
A schematic of our circuit and the corresponding physical realization are illustrated in Fig.~\ref{fig:1}. Three identical lumped LC resonators of frequency $\nu_F$ are capacitively coupled to each other in a chain to form our multimode filter. Two flux-tunable transmon\cite{koch:2007} qubits ($\nu_{Q} \approx 1-9$GHz) are capacitively coupled to the resonators at the end of the filters. For qubit frequencies $\nu_{Q,1}$, $\nu_{Q,2}$, the qubit-filter system (for $n$-modes) is described by the Hamiltonian
\begin{eqnarray}
\hat{H} & = & \hat{H}_{Q} + \hat{H}_{F} + \hat{H}_{Q-F}  \label{eqn:ham1} \\
\hat{H}_{Q} & = & h \nu_{Q,1}\hat{\sigma}^{Z}_{1}/2+ h \nu_{Q,2}\hat{\sigma}^{Z}_{2}/2 \\
\hat{H}_{F} & = & \sum_{i=1}^{n} h \nu_{F} \hat{a}^{\dagger}_{i} \hat{a}_{i} + \sum_{i=2}^{n} h g_F (\hat{a}^{\dagger}_{i} \hat{a}_{i-1} + \hat{a}^{\dagger}_{i-1} \hat{a}_{i}) \\
\hat{H}_{Q-F} & = &  h g_{Q1,F}(\hat{a}_{1}^{\dagger} \hat{\sigma}_{1}^{-}+\hat{a}_{1} \hat{\sigma}_{1}^{+}) + \nonumber \\
& & h g_{Q2,F}(\hat{a}_{n}^{\dagger} \hat{\sigma}_{2}^{-}+\hat{a}_{n} \hat{\sigma}_{2}^{+})  
\end{eqnarray}
where $\hat{a}^{\dagger}_i$ creates a photon in the $i^{\mathrm{th}}$ resonator, $\hat{\sigma}^{+(-)}$ is the raising (lowering) operator for the qubit, $\hat{\sigma}^{Z}$ is the z-Pauli operator, $g_{F}$ is the filter-filter coupling and $g_{Q,F}$ is the qubit-filter coupling.  \\
Strong coupling between the bare filter resonators splits the three degenerate resonators into three ``filter'' modes with frequencies $\nu_1,\nu_2,\nu_3=\nu_F - \sqrt{2}g_{F},\nu_F, \nu_F + \sqrt{2}g_{F}$. Each of these filter modes are a superposition of photons in the bare resonators. Crucially, every filter mode has non-zero weight in the resonators at either end of the chain so that filter photons in mode $i$ strongly couple to qubit 1 (qubit 2) with coupling $g_{Q1,Fi}(g_{Q2,Fi})$ -- this realizes our multimode strong coupling architecture. We fit the spectroscopy data in Fig.~\ref{fig:1} (c) to extract bare qubit-filter parameters $\nu_F=7.169$GHz, $g_F=118$MHz, and $g_{Q1,F}$($g_{Q2,F}$)$=135 (144)$MHz corresponding to multimode parameters $\nu_{1}=7.002$GHz, $\nu_{2}=7.169$GHz, $\nu_{3}=7.336$GHz and $g_{Q1,F2}(g_{Q2,F2})=95(102)$MHz ($g_{Q,F1}=g_{Q,F3}=g_{Q,F2}/\sqrt{2}$).\\
When the qubits are detuned from all the filter modes and the filter is empty (analogous to the stop band of a classical filter) residual interactions are mediated by virtual photons through all modes and we can rewrite Eqn.~\ref{eqn:ham1} as
\begin{equation}
\hat{H} = \hat{H}_{Q} + h J \left(\hat{\sigma}^{+}_{1} \otimes \hat{\sigma}^{-}_{2} + \hat{\sigma}^{-}_{1} \otimes \hat{\sigma}^{+}_{2}\right) + h \xi \hat{\sigma}^{Z}_{1} \otimes \hat{\sigma}^{Z}_{2}, \label{eqn:ham2}
\end{equation} 
where $J$ is the exchange term and $\xi$ is the controlled-phase (c-phase) rate. If we consider identical qubit 1 and qubit 2 filter couplings $g_Q$ and let $\Delta$ be the averaged detuning of the qubit from the bare filter mode (i.e., $\Delta=(\nu_{Q1}+\nu_{Q2}-2\nu_F)/2$), then we can approximate $J$ and $\xi$ (for an $n$-mode filter) as
\begin{eqnarray}
J & \approx & \frac{g_{Q}^2}{g_F} \left(\frac{g_F}{\Delta}\right)^{n} ,\label{eqn:J} \\
\xi & \approx & \frac{4n J^2}{\Delta}. \label{eqn:Z}
\end{eqnarray}
Notably, these rates are suppressed exponentially in the number of filter modes $n$, in terms of the small parameter $g_F/\Delta$. This is a result of destructive interference between multiple filter modes which sum coherently and enables the ability to turn off interactions with high constrast by detuning from the filter. To confirm the off-rate scaling predicted by Eqns.~\ref{eqn:J}~and~\ref{eqn:Z}, we directly measure the exchange term $J$ from qubit spectroscopy, and numerically calculate the c-phase rate. The data plotted in Fig.~\ref{fig:2} agrees well to the model with no free parameters, demonstrating the essential scaling of the multimode off-rate, and implying an off-rate less than 10kHz for a qubit-qubit detuning of 50MHz. \\
\begin{figure}[t!]
\includegraphics[width=0.45\textwidth]{\figfolder{2}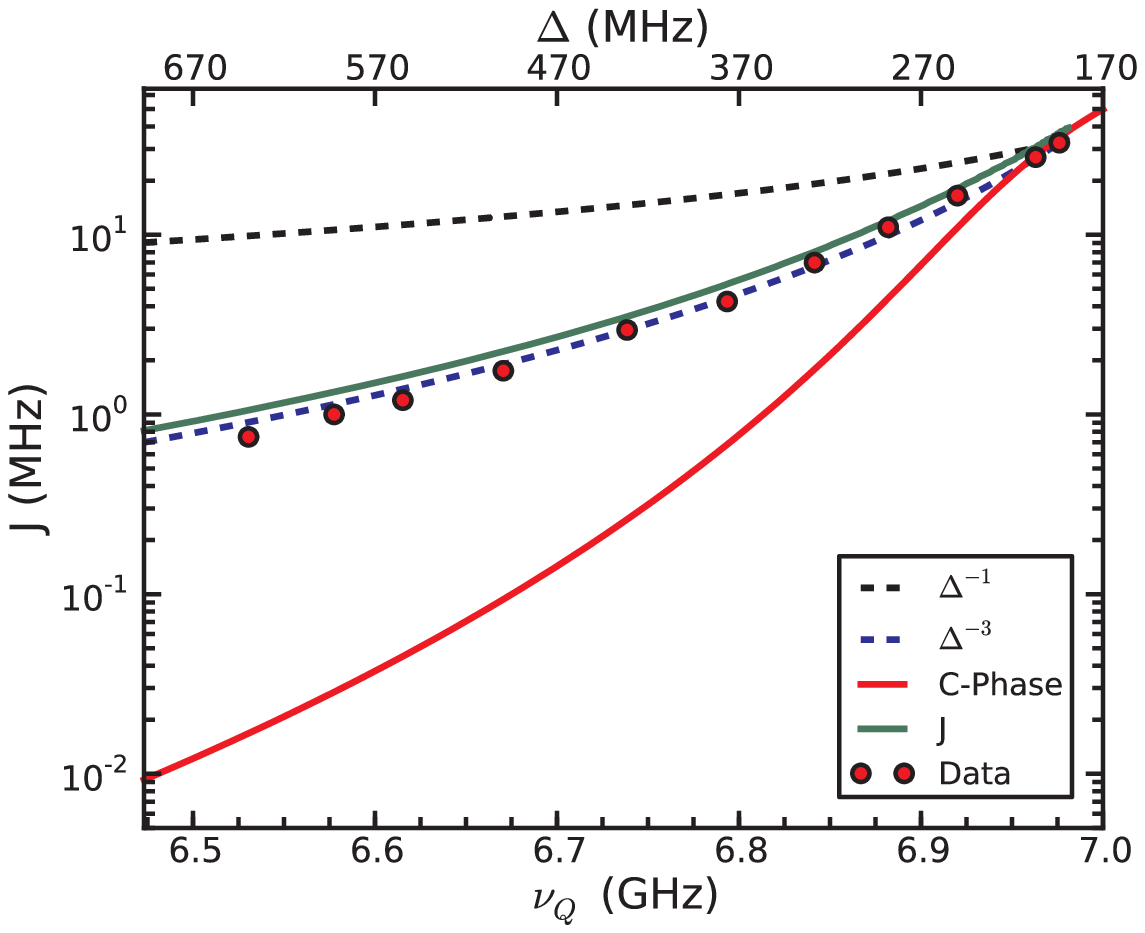}
\caption{(Color Online) {\bf Off-resonant coupling.} Qubit-qubit exchange rate $J$ as a function of the qubit frequency $\nu_Q$ (top axis, detuning $\Delta$ from the bare cavity frequency) for two different scaling laws (dashed lines), by numerical diagonalizing Eqn.~\ref{eqn:ham1} (green line), and by measuring the exchange splitting (data points). To measure the exchange splitting we analyze the qubit spectroscopy as the flux bias is tuned so that the two qubits cross in frequency, sample data is shown in Fig.~\ref{fig:1}. The splitting at the avoided crossing measures $2J$. We also plot a numerical calculation of the c-phase rate (from Eqn.~\ref{eqn:ham1}) versus the qubit 1 frequency using the filter parameters determined by the fit in Fig.~\ref{fig:1} where qubit 2 is detuned below qubit 1 by 50 MHz (red line).  \label{fig:2}}
\end{figure}
To enable strong interactions in the multimode architecture we tune the qubit frequency into resonance. In this limit, the expressions for the suppressed couplings given by Eqns.~\ref{eqn:J}~and~\ref{eqn:Z} are invalid because the virtual photon paths no longer destructively interfere. Since the qubit interacts primarily with the closest mode, the coupling strength is of order $g_Q$ (the qubit-filter coupling). Interestingly, if there is a real photon in one of the filter modes the interference is also imbalanced, hence the Stark shift is not suppressed. In both cases, the qubits interact predominantly through a single mode and so the strong interaction physics is essentially that of two qubits interacting through a single cavity.  \\
For our controlled-Z gate, we utilize these strong interactions by loading a real photon into the lowest filter mode and then employing a state-dependent one-photon Stark shift. Loading a single photon requires adiabatically traversing the qubit-filter avoided crossing shown in Fig~\ref{fig:1}, so we first study the dynamics of this crossing by performing the experiment illustrated in Fig.~\ref{fig:3} (a). We excite qubit 1, raise the qubit energy quasi-linearly through the filter in time $t$ (the flux is ramped linearly), hold for time $T-2t$, ramp back in time $t$, and then measure the excited state population. The total time $T$ is fixed to 110ns (see flux diagram inset). Because we traverse avoided crossings twice, we observe interference fringes. There are two types of fringes in Fig 3(a): fast fringes at short times and slower fringes dominant at longer times. The fast fringes correspond to ramp speeds larger than the total filter bandwidth ($\gtrsim 400$MHz) where a significant fraction of the excitation remains with the qubit\cite{kayanuma:1985}. The slower fringes correspond to the excitation being distributed over multiple filter modes and the fringe frequency is fixed by the filter mode splitting. The multimode nature of the crossing is advantageous; although the ramp is not adiabatic with the lowest filter mode unless it is slower than $\approx$25ns, the excitation remains in the filter for ramps $>$5ns. We exploit this multimode Landau-Zener physics to transfer population to the filter faster than the single mode adiabatic limit.  \\
Next, we measure the Stark shift between a single photon and a qubit in the ground state by performing a Ramsey experiment on the photon while varying the length of the interaction at different detunings as illustrated by Fig.~\ref{fig:3} (b). First, we prepare qubit 1 in the superposition state $(|g\rangle+|e\rangle)/\sqrt{2}$ and then raise the qubit frequency through the filter to create a photon superposition state. Next, we raise the frequency of qubit 2 to $\nu_{Q2,f}$ for a variable time $\tau$. After a fixed total time, we retrieve the photon from the filter, apply a $\frac{\pi}{2}$ pulse and measure the state of qubit 1. Because of the variable time interaction with qubit 2, we measure a Ramsey fringe versus $\tau$. The frequency of the fringe is the Stark shift; sample data for one of the points is shown in the inset to Fig.~\ref{fig:3}(b). Approaching the filter from below, the Stark shift increases as $\approx 1/\Delta$, and then saturates at the maximum interaction (approximately the filter splitting $\sqrt{2}g_F = 167$MHz) as qubit 2 is brought through the filter. The data agree very well with a theory curve with no free parameters, thus validating that we are loading a single photon into the lowest filter mode and that we can generate strong interactions between a qubit and a single photon.\\
\begin{figure}[htb!]
\includegraphics[width=0.44\textwidth]{\figfolder{3}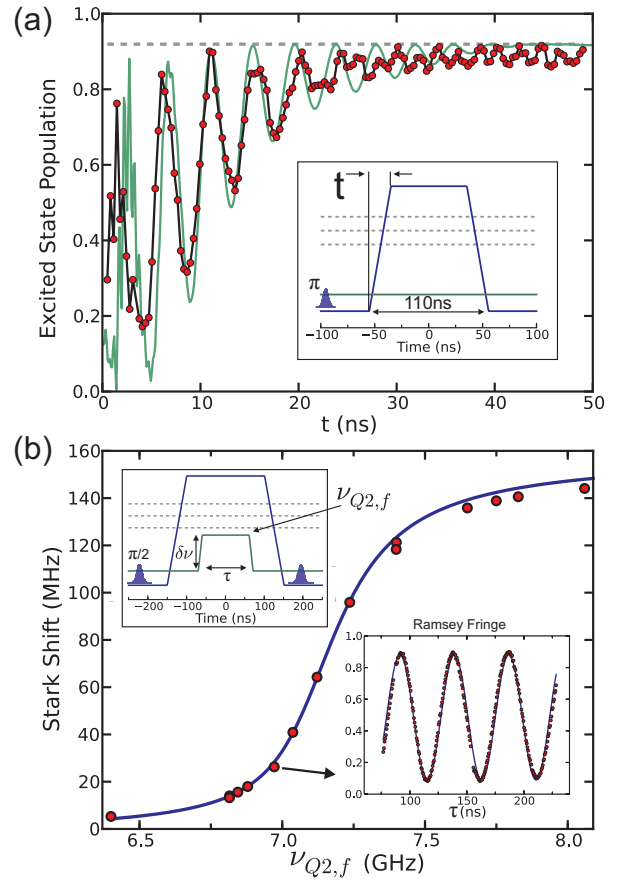}
\caption{(Color Online) {\bf Single photon loading and Stark shift.} (a) To study the dynamics of loading single photons into the filter we traverse the qubit-filter avoided crossing in variable time $t$ (protocol illustrated in inset and described in the main text). We plot the qubit excited state population versus the ramp time $t$: the solid black line is a guide to the eye, the dashed grey line is the expected maximum state population given $T_1$ decay, and the green solid line is a numerical solution of the Schrodinger equation using the Hamiltonian given by Eqn.~\ref{eqn:ham1} and scaled by $T_1$ decay. (b) To measure the Stark shift between a single photon in the lowest mode of the filter and a qubit at bare frequency $\nu_{Q2,f}$ we perform a Ramsey experiment (protocol illustrated in inset and described in the main text). We plot the Stark shift as a function of $\nu_{Q2,f}$ and compare against a theory curve (blue solid line) with no free parameters. We use $\nu_{Q2,f}=$5.3GHz as the reference height, and so set the Stark shift at that point to zero (the true Stark shift is referenced to $\delta \nu \rightarrow -\infty$). \label{fig:3}}
\end{figure}
Finally, we combine the capabilities probed in the previous two experiments --- loading a single photon into the filter and generating a strong Stark shift --- to construct a quantum logic gate. The protocol for the gate is illustrated in Fig.~\ref{fig:4}(a). First, we convert the qubit 1 excitation into a photon, then we move qubit 2 close to the filter to acquire a state dependent Stark shift, and then we return the photon back to qubit 1. While the qubit energies cross during these ramps, we observe no evidence of an exchange process since our multimode filter strongly surpresses the off-resonance interaction (Eqn.~\ref{eqn:J}). We realize a controlled-Z gate (CZ) because the conditional phase $\phi_{\mathrm{c-phase}}=\phi_{|ee\rangle}+\phi_{|gg\rangle}-(\phi_{|eg\rangle}+\phi_{|ge\rangle})$ (calculated in Fig.~\ref{fig:4}~(b)) is $\pi$. The qubits will also acquire trivial single qubit phases which we calibrate out by fine-tuning the flux pulse shape. Although the largest one-photon Stark shift occurs when we bring qubit 2 through the filter (Fig.~\ref{fig:3} (b)), the largest state-dependent interaction occurs when we bring qubit 2 just below the filter because of the full state structure of the transmon (see supplementary information). The flux pulse sequence for our CZ gate is illustrated in Fig.~\ref{fig:4}. The total gate time, 95ns, is optimized to maximize gate fidelity. For 50MHz detuning between the qubits, this implies a gate contrast (on/off rate) greater than 1000 even for relatively small $\Delta/g_F\approx 5$.\\
\begin{figure*}[htb!]
\includegraphics[width=0.9\textwidth]{\figfolder{4}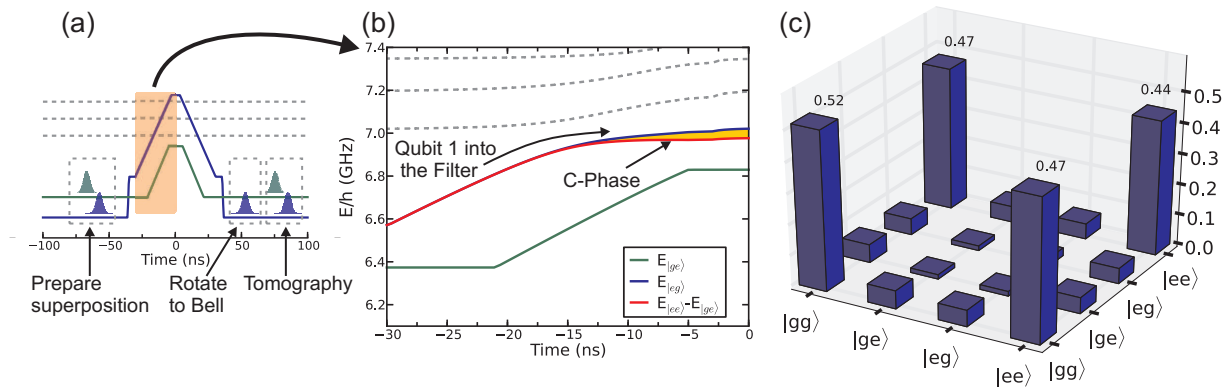}
\caption{ (Color Online) {\bf Bell state and CZ gate.} To create a Bell state between the two qubits we realize a CZ gate by using the state-dependent Stark shift between a qubit and a photon. (a) Timing diagram for our Bell state experiment illustrating the flux pulse (solid lines) used to tune the qubit frequencies and microwave pulses (Gaussian) used to perform single qubit rotations. All microwave pulses are applied to the qubits simultaneously, but are offset in the diagram for clarity. (b) The energy levels calculated from Eqn.~\ref{eqn:ham1} in the orange region highlighted in (a). When qubit 1 approaches the filter there is an avoided crossing with the first filter mode which converts the qubit excitation into a filter photon. When qubit 2 is raised, the energy of the filter photon depends on the state of qubit 2, which generates a c-phase. The total phase is the yellow area indicated in the graph. (c) Absolute value of the density matrix elements after state tomography of the Bell state produced by the gate. Details of the state tomography are given in the supplementary information and fidelities are discussed in the main text.  \label{fig:4}}
\end{figure*}
To demonstrate the gate we prepare a Bell state by applying single qubit pulses before and after the gate. Ideally this process creates the Bell state $|\Psi_{Bell}\rangle=(|gg\rangle+e^{i\Phi} |ee\rangle)/\sqrt{2}$. To characterize the expected density matrix we perform state tomography\cite{james:2001} on both qubits after the gate (see Fig.~\ref{fig:4} (c)). The fidelity $F=\langle \Psi_{Bell} | \rho_{meas} | \Psi_{Bell}\rangle$ is 0.947$\pm0.005_{\mathrm{stat}}\pm 0.01_{\mathrm{sys}}$ corresponding to concurrence of 0.926$\pm0.01_{\mathrm{stat}}\pm 0.02_{\mathrm{sys}}$\cite{wootters:1998}. We also measure a full process fidelity of 0.89 (errors and tomography details are discussed in the supplementary information). Our fidelity is comparable to other contemporary results (two-qubit entangled states have been produced with state fidelities up to 99.5\%\cite{barends:2014} and concurrence of 0.994 \cite{chow:2012}), and is limited by our lifetime, rather than the protocol. One advantage of our protocol is that our gate is relatively insensitive to inhomogeneous broadening due to flux noise; once the qubit excitation is a photon in the filter, the energy is not flux dependent. Several improvements are possible, for example, engineering a flux insensitive bias point below the filter for state preparation \cite{strand:2013}, utilizing new materials \cite{chang:2013} and material processing for high Q resonators \cite{megrant:2012}, as well as reducing the total gate time using techniques from optimal control for crossing the filter.\\
In conclusion, we have demonstrated a new multimode architecture for coupling superconducting qubits. We measured that the off-resonance coupling is suppressed exponentially in the number of modes, while still maintaining strong interactions when the qubits are tuned close to resonance. We used these capabilities to realize a high-contrast controlled-Z gate. Further, this work indicates a need to develop a microwave filter theory for coherent quantum systems. The multimode architecture is a promising platform for realizing lattice based quantum simulations and photonic registers for quantum information processing. \\
\textbf{Acknowledgments}: We acknowledge support from the University of Chicago MRSEC, Army Research Office under grant W911NF-12-1-0608, the Alfred P. Sloan Foundation, NSF under grant DMR-1151839, and DARPA grant N66001-11-1-4123. We acknowledge David Czaplewski at the Argonne Center for Nanomaterials for assistance with the optical lithography. We thank J. Chow and B. Johnson for discussions and J. Thywissen for manuscript comments. \\

\end{document}